\documentclass{article}
\usepackage{spconf,amsmath,graphicx}
\usepackage{url}
\usepackage{amsfonts}
\usepackage{booktabs}
\usepackage{siunitx}
\usepackage{multirow}
\usepackage{enumitem}
\usepackage[numbers]{natbib}
\usepackage[symbol]{footmisc}

\usepackage{hyperref}
\hypersetup{colorlinks,bookmarks=false}

\usepackage{color}

\usepackage{enumitem}

\title{VocBench: A Neural Vocoder Benchmark for Speech Synthesis}
\name{Ehab A. AlBadawy\sthanks{Work done during an internship at Facebook AI.}$^{\mathsection}$,
Andrew Gibiansky$^{\dagger}$,
Qing He$^{\dagger}$,
Jilong Wu$^{\dagger}$,
Ming-Ching Chang$^{\mathsection}$,
Siwei Lyu$^{\ddagger}$}
  
\address{$^{\dagger}$Facebook AI, USA\\
$^{\mathsection}$ University at Albany, State University of New York, USA\\
$^{\ddagger}$University at Buffalo, State University of New York, USA}
\begin{document}
\setcitestyle{square}
%
\maketitle
%

\begin{abstract}

Neural vocoders, used for converting the spectral representations of an audio signal to the waveforms, are a commonly used component in speech synthesis pipelines.
It focuses on synthesizing waveforms from low-dimensional representation, such as Mel-Spectrograms.
In recent years, different approaches have been introduced to develop such vocoders.
However, it becomes more challenging to assess these new vocoders and compare their performance to previous ones.
To address this problem, we present VocBench, a framework that benchmark the performance of state-of-the art neural vocoders.
VocBench uses a systematic study to evaluate different neural vocoders in a shared environment that enables a fair comparison between them.
In our experiments, we use the same setup for datasets, training pipeline, and evaluation metrics for all neural vocoders.
We perform a subjective and objective evaluation to compare the performance of each vocoder along a different axis.
Our results demonstrate that the framework is capable of showing the competitive efficacy and the quality of the synthesized samples for each vocoder.
VocBench framework is available at \href{https://github.com/facebookresearch/vocoder-benchmark}{https://github.com/facebookresearch/vocoder-benchmark}.
\end{abstract}
\noindent\noindent \textbf{Index Terms}: Speech, Self-Supervised Learning, Model Generalization, Benchmark, Evaluation
\vspace{-0.4cm}


\section{Introduction}
\vspace{-0.2cm}

\label{sec.intro}

Through out the years, speech synthesis techniques have gone through different phases of improvements, from knowledge-based approaches~\cite{tatham1985integrated, grigoras1999fuzzy, anumanchipalli2010klattstat} to data-based ones~\cite{tokuda2000speech, wang2017tacotron, li2019neural}. To date, there are two types of speech synthesis algorithms, text to speech, which converts input text to audio signals, and voice conversion, which transforms an input audio to different identities or styles.
Regardless this difference, most of the recent speech synthesis approaches \cite{shen2018natural, kameoka2018stargan,AlBadawy2020} rely on \emph{neural vocoders} to generate the final waveform for more natural sounding speech synthesis. 

In this context, a vocoder is designed to synthesize waveform from the frequency acoustic features such as Mel-Spectrograms or F0 frequencies. 
For many years, the state-of-the art (SOTA) methods were DSP-based approaches~\cite{griffin1984signal} for vocoder development. 
While the advantage of fast speech generation time, the quality of the synthesized waveform is largely limited due to the assumptions under the heuristics. 
In the recent years, more sophisticated vocoders have been developed based on the use of deep neural networks for more enhanced quality for the generated speech. These methods include autoregressive approaches~\cite{vanwavenet,kalchbrenner2018efficient}, Generative Adversarial Networks (GANs) approaches~\cite{kumar2019melgan,yang2021multi,yamamoto2020parallel}, and diffusion based approaches~\cite{chen2020wavegrad,kong2020diffwave}.
Due to the different variables in the evaluation process; datasets selection, hardware configuration, or the evaluation metrics used, how best to compare and evaluate these different approaches remains an open challenge.


\begin{figure}[t]

\begin{minipage}[b]{1.0\linewidth}
\centerline{
\includegraphics[width=1\textwidth]{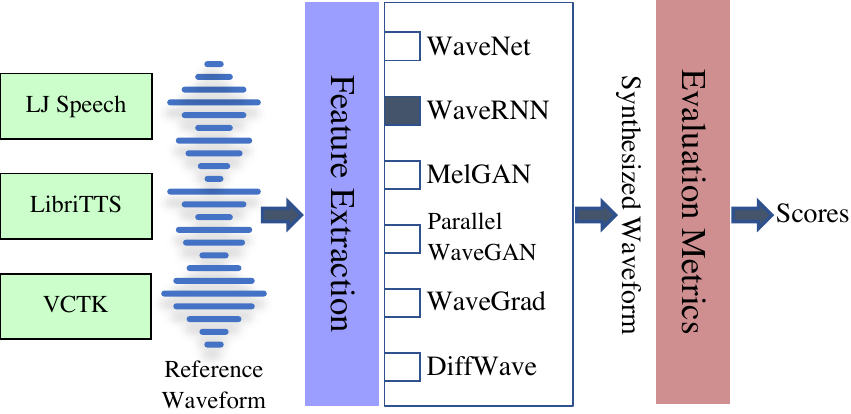}
\vspace{-0.2cm}
}
\end{minipage}
\caption{An overview of the proposed VocBench
framework.}
\label{fig:vocbench_pipeline}
\vspace{-7mm}
\end{figure}

In this work, we present VocBench framework, a comprehensive benchmark for vocoder quality and speed evaluations.
More specifically, we build VocBench to train and test neural vocoders in a shared environment with public datasets.
We construct three datasets, including one single-speaker and two multi-speaker scenarios, and then we train six vocoders covering three different categories; Autoregressive, GAN, and Diffusion approaches. All of the vocoders Follow the same training and evaluation pipeline.
We design the following two main experiments. 
First, we test the efficacy of each vocoder in synthesizing the waveform from lower-dimensional features such as Mel-Spectrogram.
Second, we test the generalizability of each vocoder in synthesizing speech for speakers who are not included in the training set. Figure~\ref{fig:vocbench_pipeline} provides an overview of the proposed framework.

Recently, various studies have been conducted for neural vocoder evaluation. 
Govalkar {\em et al.}~\cite{govalkar2019comparison} conducted a study with six autoregressive-based vocoders and two additional phase reconstructions vocoders. 
Airaksinen {\em et al.}~\cite{airaksinen2018comparison} adopted classical methods for vocoder design in their study. Both of these works used MUSHRA~\cite{schoeffler2015towards} as their main evaluation metric to compare the performance of each vocoder.
In this study, we extend the vocoder implementations to include both GAN and diffusion based models.
Additionally, we carry on the evaluation using both subjective and objective metrics.
We use the Mean Opinion Score (MOS) test as a subjective evaluation.
To evaluate each of the different vocoders objectively, we used the following four different evaluation metrics:
Structural Similarity Index Measure (SSIM) \cite{wang2004ssim}, Fréchet Audio Distance (FAD) \cite{kilgour2018fr}, Log-mel Spectrogram Mean Squared Error (LS-MSE), and Peak Signal-to-Noise Ratio (PSNR). More details about the experiment setup and evaluation metrics are presented in $\S$~\ref{sec.exp}.

%


 \vspace{-0.35cm}
\section{Neural Vocoders}
\label{sec.vocoders}
\vspace{-0.2cm}

In this section we describe the three main categories of the neural vocoders we use in our study, namely, the autoregressive models ($\S$~\ref{sec:autoregressive}),  GAN-based models ($\S$~\ref{sec:GAN}), and diffusion models ($\S$~\ref{sec:diffusion}).


\vspace{-0.5cm}
\subsection{Autoregressive Models}
\label{sec:autoregressive}
\vspace{-0.2cm}

The key feature of the autoregressive models is that they are designed as probabilistic models to predict the probability of each waveform sample based on the previous samples. 
This allows to generate a high quality and natural speech signal. 
However, due to the sample-by-sample generation process, the overall synthesis speed is slow compared to other methods. In the following, we will consider two main autoregressive models: WaveNet and WaveRNN.

The \textbf{WaveNet}~\cite{vanwavenet} model works on the waveform level to achieve long-range temporal dependency through the depth of the model. It combines a stack of causal filters and dilated convolutions to help their receptive fields grow exponentially with depth. We use the open-source implementation from \cite{ryuichi_yamamoto_2018_1472609} with different configurations of input types and loss functions. More details are provided in $\S$~\ref{sec.exp}.

The autoregressive {\bf WaveRNN}~\cite{kalchbrenner2018efficient} architecture utilizes a recurrent neural network (RNN) for sequential modeling of the target waveform. A single layer RNN with a dual softmax layer is used. 


\vspace{-0.5cm}
\subsection{GAN Models}
\label{sec:GAN}
\vspace{-0.2cm}

GAN-based vocoders have shown remarkable performance often exceeding autoregressive models in the speed and quality of the synthesized speech. They utilize the main idea of GANs~\cite{goodfellow2014generative}, and use a {\em generator} to model the waveform signal in the time domain as well as a {\em discriminator} to assist the quality of the generated speech. Different variants of the GAN-based vocoders have been introduced. 
We consider two representative models in the following: MelGAN and Parallel WaveGAN.

The \textbf{MelGAN}~\cite{kumar2019melgan} takes the standard GAN as the network architecture for fast waveform generation. A fully convolutional model is used for high quality Mel-Spectrogram inversion. With fewer number of parameters compared to autoregressive model, MelGAN achieves higher real-time factor on both GPU and CPU without the need of hardware specific optimization.


The \textbf{Parallel WaveGAN}~\cite{yamamoto2020parallel} architecture is another GAN based model that is distillation-free, fast, and small-footprint for waveform synthesis. Parallel WaveGAN jointly optimizes the waveform-domain adversarial loss and multi-resolution short-time Fourier transform (STFT) loss. 

\vspace{-0.5cm}
\subsection{Diffusion Models}
\label{sec:diffusion}
\vspace{-0.2cm}

Diffusion probabilistic models are another class of generative models entailing two main processes: the diffusion process and the reverse process~\cite{ho2020denoising}.
The diffusion process is defined as a Markov chain process that gradually add Gaussian noise to the original signal until it gets destroyed.
The reverse process on the other hand is a denoising process which progressively removes the added the Gaussian noise and restores the original signal.
We included two diffusion-based vocoders in our study: WaveGrad and DiffWave.

The \textbf{WaveGrad}~\cite{chen2020wavegrad} model architecture is built on prior work on score matching~\cite{vincent2011connection} and diffusion probabilistic models~\cite{ho2020denoising}. The WaveGrad model takes a white Gaussian noise as input, and conditioning on the Mel-Spectrogram to iteratively refine the signal via a gradient-based sampler.

\textbf{DiffWave}~\cite{kong2020diffwave} is a versatile diffusion probabilistic model for waveform synthesis that works well under both conditional and unconditional scenarios. 
Using a white Gaussian noise as input, DiffWave performs Markov chain process with a constant number of steps to gradually generate a structured waveform~\cite{sohl2015deep, goyal2017variational, ho2020denoising}. The model is trained to optimize a variant of the variational bound on the data likelihood.

 \vspace{-0.2cm}
\begin{table*}[t]
\small
\caption{Evaluation results for the four objective metrics (SSIM, LS-MSE, PSNR, and FAD) and the 5-scale MOS with 95\% confidence intervals evaluated on the three datasets: LJ Speech, LibriTTS and VCTK. We welcome researchers to submit or update their results to our GitHub repository for comparisons.
}
\vspace{-.3cm}
\label{tab:obj-subj-eval}
\begin{center}
\setlength{\tabcolsep}{3pt}
\begin{tabular}{c|c|cc|cc|cc|cc}\toprule

\multirow{2}{*}{\textbf{Metric}}
&\multirow{2}{*}{\textbf{Corpus}} 
&\multirow{2}{*}{\textbf{WaveNet}}
&\multirow{2}{*}{\textbf{WaveRNN}}  
&\multirow{2}{*}{\textbf{MelGAN}}
&\multirow{1.2}{*}{\textbf{Parallel}}
&\multirow{2}{*}{\textbf{WaveGrad}}
&\multirow{2}{*}{\textbf{DiffWave}}    
&\multirow{2}{*}{\textbf{Griffin-Lim}}
&\multirow{1.2}{*}{\textbf{Ground}}
\\&&&&&\multirow{1.2}{*}{\textbf{WaveGAN}}&&&&\multirow{1.2}{*}{\textbf{Truth}}
\\ \midrule

\multirow{3}{*}{SSIM}   &LJ Speech & 0.66 & 0.62 & 0.89 & 0.84 & 0.76 & 0.82 & 0.90 & - \\
                        &LibriTTS & 0.056 & 0.53 & 0.91 & 0.86 & 0.71 & 0.74 & 0.89 & - \\
                        &VCTK & 0.46 & 0.43 & 0.88 & 0.79 & 0.59 & 0.64 & 0.86 & - \\
\midrule  
\multirow{3}{*}{LS-MSE}   &LJ Speech & 0.006 & 0.010 & 0.001 & 0.002 & 0.006 & 0.006 & 0.001 & - \\
                        &LibriTTS & 0.008 & 0.008 & 0.001 & 0.001 & 0.005 & 0.006 & 0.001 & - \\
                        &VCTK  & 0.009 & 0.010 & 0.001 & 0.002 & 0.007 & 0.007 & 0.001 & - \\
\midrule  
\multirow{3}{*}{PSNR}   &LJ Speech & 23.20 & 20.36 & 28.53 & 26.70 & 22.57 & 22.51 & 28.77 & - \\
                        &LibriTTS & 21.54 & 21.17 & 29.98 & 28.62 & 22.94 & 22.18 & 29.03 & - \\
                        &VCTK & 21.36 & 20.40 & 30.40 & 28.17 & 21.54 & 21.22 & 28.77 & - \\
\midrule  
\multirow{3}{*}{FAD}   &LJ Speech & 1.05 & 3.43 & 1.51 & \textbf{0.92} & 3.12 & 3.62 & 2.69 & 0.31 \\
                        &LibriTTS & 1.55 & 2.60 & 2.95 & \textbf{1.41} & 3.10 & 3.74 & 4.27 & 1.23 \\
                        &VCTK & \textbf{0.99} & 3.59 & 1.76 & 1.22 & 4.10 & 5.59 & 3.92 & 0.61 \\  
\midrule                          
\multirow{3}{*}{MOS}   &LJ Speech & 3.68$\pm$0.037& 3.96$\pm$0.089& 3.73$\pm$0.075& 3.99$\pm$0.059& 3.85$\pm$0.068& \textbf{4.07$\pm$0.060}& 3.68$\pm$0.082& 4.10$\pm$0.059\\
                        &LibriTTS & 3.75$\pm$0.107 & 3.74$\pm$0.099 & 3.50$\pm$0.086 & \textbf{3.82$\pm$0.069} & 3.48$\pm$0.083 & 3.80$\pm$0.073 & 3.36$\pm$0.092 & 4.03$\pm$0.065 \\
                        &VCTK & \textbf{3.95$\pm$0.032} & 3.94$\pm$0.089 & 3.75$\pm$0.074 & 3.87$\pm$0.068 & 3.77$\pm$0.074 & 3.86$\pm$0.069 & 3.66$\pm$0.079 & 3.98$\pm$0.064 \\
\bottomrule
\end{tabular}
\end{center}
\vspace{-9mm}
\end{table*}

\vspace{-0.2cm}
\section{Dataset and Experiments}
\label{sec.exp}
\vspace{-0.2cm}

\subsection{Dataset and Feature Extraction}
\label{sec.exp.dataset}
\vspace{-0.1cm}

We use three datasets in this study: LJ Speech for single-speaker scenario as well as LibriTTS and VCTK for multi-speaker scenarios. For all of the three different datasets, the train, validation, and test splits are fixed across the different vocoders we use in our study.

The \textbf{LJ Speech} dataset~\cite{ljspeech17} consists of $13,100$ short audio clips of a single speaker reading passages from $7$ non-fiction books. A transcription is provided for each clip. The length of each clip varies from $1$ to $10$ seconds, and the total length is approximately $24$ hours. We reserve the first $20$ clips for testing, and the following $10$ clips for validation. The rest of the clips are used for training.

The \textbf{LibriTTS} dataset~\cite{zen2019libritts} is a multi-speaker English corpus of approximately $585$ hours of read English speech at $24$kHz sampling rate. It is derived from the original materials (mp3 audio files from LibriVox and text files from Project Gutenberg) of the LibriSpeech corpus. We use \emph{train-clean-$100$} and \emph{train-clean-$360$} subsets for training with about $1150$ speakers and $25$ minutes of recordings on average per speaker. For validation and test splits, we use the \emph{dev-clean} and \emph{test-clean} subsets respectively.

The \textbf{VCTK} corpus~\cite{Veaux2017CSTRVC} includes speech data uttered by 110 English speakers with various accents. Each speaker reads out about 400 sentences selected from a newspaper. We randomly select $85\%$ of the samples for training data, $10\%$ for validation, and $5\%$ for testing.

\textbf{Log-spectrogram computations.} The speech signals in the three datasets are re-sampled to $24$ kHz. We extract the $80$-dimensional Mel-Spectrogram features using 40 ms Hanning window, $12.5$ ms frame shift, $1024$-point FFT, and $0$ Hz \& $12$ kHz lower \& upper frequency cutoffs. We then perform log dynamic range compression on the resulting Mel-Spectrogram features followed by a min-max normalization.

\vspace{-0.5cm}
\subsection{Training Setup}
\label{sec.exp.config}
\vspace{-0.1cm}


For the training each of the vocoders in our study, we conduct a hyperparameter search and report the best model configuration on the three different datasets described in $\S$~\ref{sec.exp.dataset}. We implement our framework based on the PyTorch library, and training is performed on a Tesla V100 GPU. For reproducibility, we use the Amazon Web Services (AWS) to compute the evaluation metrics. More specifically, we use c5.4xlarge AWS instance for the CPU computations; it has 16 vCPU of Intel Xeon Processors with 3.6GHz frequency. For the GPU computations we use p3.2xlarge AWS instance, it has 8 vCPU of Intel Xeon Processors with 2.3GHz frequency and one NVIDIA Tesla V100 GPU. 

For each of the vocoders, we start from the original configuration provided in the respective open-source implementation. However for WaveNet, there are different configurations that we can chose from. They vary in terms of input types and loss functions that can be used. For the input, we can use either raw waveform or pre-processed waveform using $\mu$-law compression. For the loss function, there are two different options we can choose from: Mixture of Logistics (MoL-loss) and a single Gaussian distribution ({\em normal-loss}). We run different versions of the WaveNet model using each of the possible configurations and report the one with the best performance. We found that on LJ Speech and VCTK, it is better to use $\mu$-law compression on the input waveform, while for LibriTTS, raw waveform input achieves the best results. For the loss function, using {\em normal-loss} helps to increase the overall performance.

\vspace{-0.4cm}
\subsection{Evaluation}
\label{sec.exp.eval}
\vspace{-0.1cm}

It is crucial what metrics we used for evaluation. The core idea is to evaluate multiple vocoders along different axes in both numerical and qualitative aspects. We adopt the following metrics is this study:
\vspace{-.1cm}
\begin{itemize}[leftmargin=*]\itemsep -.2em
\item \textbf{Mean Opinion Score (MOS)} is a subjective numerical measure of the human-judged overall quality after listening to a sample. We use 5-scale MOS to assess the quality for the synthesized speech sample. We report the MOS for each vocoder as well as the ground truth over 10 samples from the test set.

\item \textbf{Structural Similarity Index Measure (SSIM)}~\cite{wang2004ssim} is a quantitative metric that measures the similarity between two given images. We perform SSIM in the frequency domain to compare the synthetic spectrogram with the real-world sample.

\item \textbf{Fréchet Audio Distance (FAD)} \cite{kilgour2018fr} measures the quality and diversity of the generated samples. FAD score is the distance between two multivariate Gaussian estimated on sets of embeddings, {\em i.e.} the background and evaluation embeddings. To generate these feature embeddings, FAD use a VGG model \cite{hershey2017cnn} trained on large YouTube dataset as the audio classifier.

\item \textbf{Log-mel Spectrogram Mean Squared Error (LS-MSE)} is computed between the ground truth spectrogram sample and a generated one. We use the computation in $\S$~\ref{sec.exp.dataset} to obtain the Log-mel Spectrogram for the synthesized speech samples. The LS-MSE can be interpreted as a measure of how close the low-dimensional representation of the spectrogram is when compared to the ground truth spectrogram.

\item \textbf{Peak Signal-to-Noise Ratio (PSNR)} is the ratio of the power of a peak signal, which is the magnitude of the best-case output of a signal to the power of the noise at the peak measured in dB. We apply PSNR computation in the frequency domain, where the peak signal of the output is 1 and the distorting noise is represented by LS-MSE.
\vspace{-.5cm}

\end{itemize}

\vspace{-0.105cm}
\subsection{Results and Discussion}
\label{sec.exp.results}
\vspace{-0.1cm}

Table~\ref{tab:obj-subj-eval} shows the results of the five objective and subjective evaluation metrics we described in $\S$~\ref{sec.exp.eval}.
Each of the metrics are computed using 20 audio samples from each dataset.
For MOS, we report the mean value as well as the 95\% confidence intervals.
We use the Griffin-Lim vocoder \cite{griffin1984signal} as a baseline to compare with each of the other vocoders in our study.

FAD and MOS metrics show close correlation specially in GAN-based vocoders. both metrics have the same best performing models in each dataset except LJ Speech. MOS reports Diffwave as the best performing vocoder for LJ Speech with a MOS score of $4.07 \pm 0.06$, while Parallel WaveGAN achieves the best FAD score of $0.92$ (and the second best in terms of MOS $3.99 \pm 0.059$).
For LibriTTS dataset, Parallel WaveGAN has the best performance for both FAD ($1.41$) and MOS ($3.82 \pm 0.69$).
As for VCTK dataset, WaveNet achieves lower FAD ($0.99$) and higher MOS ($3.95 \pm 0.032$) scores compared to other vocoders.

Observe that when using FAD and MOS metrics, each of the different models achieves their best performance on LJ Speech dataset while having lower performance on VCTK and LibriTTS, respectively. 
This is due to the fact that LJ Speech is a single-speaker dataset which makes it easier to train and evaluate on. 
On the other hand, VCTK and LibriTTS are multi-speaker datasets. When LibriTTS is used for speaker generalizability, experiment results show that it makes the scenario more challenging. 

Table~\ref{tab:model-specs} shows the model complexity of each vocoder and how that affects the voice synthesis time. We compare the following aspects of the neural vocoder models: the model parameter size, the number of Floating Point Operations per Second (FLOPS) of a speech sample, total training iterations, and Real Time Factor (RTF).

The autoregressive models, WaveNet and WaveRNN have a consistent number of parameters (3.79 and 4.35 Million parameters respectively) and FLOPS (89.65 and 94.98 GFLOPS respectively), when compared to other models.
We excluded the RTF computation for the autoregressive model as they are significantly slower compared to other vocoders in our study. 
For real time applications, custom kernels are used for autoregressive models such as LPCNET \cite{valin2019lpcnet}.

For GAN-based vocoders, we report the number of parameters and FLOPS for the generator model only.
MelGAN has fewer number of FLOPS (3.01 GLOPS) when compared to the Parallel WaveGAN (31.26 GLOPS).
That difference is reflected in the RTF values where MelGAN has RTF $0.001$ RTF for GPU and 0.029 for CPU. On the other hand, Parallel WaveGAN achieves 0.002 RTF on GPU and 0.576 on CPU.

In Diffusion-based vocoders, WaveGrad has relatively higher number of parameters (15.81 Million parameters) compared to DiffWave, while both models maintain the same order of magnitude for the number of FLOPS (33.75 and 31.70 GFLOPS respectively). We report the computation of a single step of the inference process for both the number of model parameters and FLOPS. In our experiments, during inference we use 50 steps noise scheduler for WaveGrad and 6 steps for DiffWave Following the original implementation. This explains the higher RTF obtained for both vocoders in comparison to GAN-based, where WaveGrad has 0.381 RTF on GPU and 9.858 on CPU, respectively. DiffWave reports 0.070 and 4.452 RTF on GPU and CPU respectively.




 \vspace{-0.2cm}
\begin{table}[t]
\caption{Space and time complexity for vocoders under evaluation in terms of: (1) the number of parameters, (2) computation FLOPS, and (3) their corresponding RTF using on GPU and CPU setup. \#Param for GANs (MelGAN and Parallel WaveGAN) are the for the generator model only, and for a single step of inference process for the diffusion models (WaveGrad and DiffWave).
\vspace{0.2cm}
}
\label{tab:model-specs}
\centerline{
\setlength{\tabcolsep}{3pt}
\begin{tabular}{c|cccc} \toprule

\multicolumn{1}{c|}{\bf \multirow{2}{*}{Model}}  
&\multicolumn{1}{c}{\bf \multirow{2}{*}{\#Param (M)}} 
&\multicolumn{1}{c}{\bf \multirow{2}{*}{GFLOPS}}  
&\multicolumn{2}{c}{\bf RTF}  \\
&&&GPU & CPU
\\ \midrule
WaveNet             & 3.79  & 89.65 & - & - \\
WaveRNN             & 4.35  & 94.98 & - & - \\
\midrule
MelGAN              & 3.05*  & 3.01&0.001 & 0.029\\
Parallel WaveGAN    & 1.34*  & 31.26&0.002 & 0.576\\
\midrule
WaveGrad            & 15.81*  & 33.75&0.381 & 9.858\\
DiffWave            & 2.62*   & 31.70&0.070 & 4.452 \\
\bottomrule
\end{tabular}
}
\vspace{-7mm}
\end{table}





\vspace{-0.2cm}
\section{Conclusion}
\vspace{-0.3cm}

We present VocBench, a framework for general purpose benchmark of neural vocoders on the speech synthesis task.
VocBench provides the speech community a standard and comprehensive appraoch for neural vocoders evaluation.
Our results show both the objective and subjective differences for the vocoders that are included in our study.
We have open-sourced our toolkit for training and evaluating neural vocoders in GitHub.
We welcome the community to contribute and share their own implementations and evaluate them against other SOTA vocoders.


\bibliographystyle{IEEEbib}
\bibliography{strings,vocbench}

\end{document}